\documentclass[conference]{IEEEtran}
\ifCLASSINFOpdf
\else
\fi

\usepackage{geometry}
\usepackage{mathptmx}       % selects Times Roman as basic font
\usepackage{helvet}         % selects Helvetica as sans-serif font
\usepackage{courier}        % selects Courier as typewriter font
\usepackage{type1cm}        % activate if the above 3 fonts are
\usepackage{amsmath}                            % not available on your system
\usepackage{makeidx}         % allows index generation

\usepackage{graphicx}        % standard LaTeX graphics tool
\usepackage{multicol}        % used for the two-column index
\usepackage[bottom]{footmisc}% places footnotes at page bottom
\usepackage[colorlinks=true]{hyperref} %
\hypersetup{
backref=true, %permet d'ajouter des liens dans...
pagebackref=true,%...les bibliographies
hyperindex=true, %ajoute des liens dans les index.
colorlinks=true, %colorise les liens
breaklinks=true, %permet le retour à la ligne dans les liens trop longs
urlcolor= blue, %couleur des hyperliens
linkcolor= blue, %color of internal links
citecolor=blue,   %color of links to bibliography
bookmarks=true, %créé des signets pour Acrobat
bookmarksopen=true, %si les signets Acrobat sont créés,
}

\begin{document}
\title{\textbf{Energy-Aware Scheme used in Multi-level Heterogeneous Wireless Sensor Networks}}

\author{\IEEEauthorblockN{Mostafa SAADI\IEEEauthorrefmark{1}\IEEEauthorrefmark{4}, Moulay Lahcen HASNAOUI\IEEEauthorrefmark{3}, Abderrahim BENI HSSANE\IEEEauthorrefmark{4}, \\Said BENKIRANE\IEEEauthorrefmark{4}, Mohamed LAGHDIR\IEEEauthorrefmark{4}}\\

\IEEEauthorblockA{\IEEEauthorrefmark{4}MATIC Laboratory, Mathematics and Computer Science Department, Faculty of Sciences,\\
 Chouaïb Doukkali University, El Jadida, Morocco.}\\

\IEEEauthorblockA{\IEEEauthorrefmark{3}Computer Science Department, Faculty of Sciences Dhar el Mahraz, \\
Sidi Mohammed Ben Abdellah University, Fez, Marocco.}\\

\IEEEauthorblockA{saadi\_mo@yahoo.fr, mlhnet2002@yahoo.ca, abenihssane@yahoo.fr, sabenk1@hotmail.com,  laghdirm@yahoo.fr}\\

\IEEEauthorrefmark{1}Corresponding Author
}


%Telephone: (800) 555--1212, Fax: (888) 555--1212}


\maketitle

\begin{abstract}
%\boldmath
The wireless sensor networks (WSNs) is a power constrained system, since nodes run  on limited power batteries which shorten its lifespan.The main challenge facing us in the design and conception of Wireless Sensor Networks (WSNs) is to find the best way to extend their life span. The clustering algorithm is a key technique used to increase the scalability and life span of the network in general. In this paper, we propose and evaluate a distributed energy-efficient clustering algorithm for WSNs. This heterogeneous-energy protocol is a new clustering algorithm to decrease probability of failure nodes and in which we introduce the node's remaining energy  so as to  determine the cluster heads. We study the impact of heterogeneity of nodes on WSNs that are hierarchically clustered. Finally, simulation results show that the proposed algorithm increases the life span of the whole network and performs better than LEACH  and EEHC according to the metric:first node dies.\\
\end{abstract}


%\cfoot{978-1-4673-4766-2/12/\$31.00 \copyright 2012 IEEE}
%\thispagestyle{fancy}

\IEEEpeerreviewmaketitle

\section{Introduction}
\label{sec:1}
Continued enhancement of Micro-Electro-Mechanical Systems (MEMS) and wireless communication technologies have enabled the deployment of large scale wireless sensor networks (WSNs). It comprises a big number of sensor nodes deployed in ad hoc manner in an unreachable field to give the end-user the ability to instrument, observe, and react to events and phenomena in a specified environment. WSNs provide unforeseen applications: ranging from military applications such as battlefield mapping and target surveillance, to creating context-aware homes; the number of applications is endless \cite{Akyildiz,Mhatre,taruna}.

Since they are exposed to atrocious and dynamic environments and limited in their energy level,
processing power and sensing ability, WSNs must deliver only processed and concise data. Therefore,
any inefficient use of these WSNs leads to a poor performance and consequently a short life cycle.
Routing techniques are the most important issue for networks where resources are limited.\cite{Estrin,Min,Heinzelman00}

In most of the applications, sensors are supposed to spot the events and then send the collected data to the Base Station (BS) where parameters characterizing these events are evaluated. Since the cost of forwarding data is higher than computation, \cite{Akyildiz,Mhatre,Min}, clustering sensors into groups  so as to communicate information only to cluster heads which communicate information to the processing center (BS), is a kind of key technique used to reduce energy consumption and  then increase the life span of the network \cite{Heinzelman00,Heinzelman02}.

In this respect, there are two types of schemes that operate differently. The conventional centralized algorithms operate with
a global knowledge of the whole network and any error in transmission or a failure of a critical node will
potentially bring about a serious protocol failure; whereas the distributed algorithms are executed locally with
partial nodes, which can prevent any failure caused by a single node \cite{Heinzelman00,Heinzelman02,SEP,DEEC,Koulali}.

In this paper, we propose a new energy-efficient cluster head selection algorithm to reduce energy consumption dubbed EASM. This heterogeneous-energy protocol decreases the probability of failure nodes and in which we introduce the node's remaining energy  so as to  determine the future cluster heads.

The operation of this algorithm is divided into rounds. Each of these rounds consists of a set-up and a steady-state phase. During the set-up phase cluster-heads are determined and the clusters are organized. During the steady-state phase data transfers to the base station occur. This protocol is proposed to increase the whole network life span on a heterogeneous network with a BS located far away from the sensor area.

The remainder of this paper is organized as follows. Section \textbf{~\ref{sec:2}} presents the related work and describes the heterogeneous WSN model. Section \textbf{~\ref{sec:3}} exhibits the details and analyzes the properties of the newest one. Section \textbf{~\ref{sec:4}} evaluates the performance of our protocol by simulations and compares it with other existing protocols. Finally, Section \textbf{~\ref{sec:5}} gives concluding remarks.

\section{Problem out line }
\label{sec:2}
\subsection{Related Work }
In most WSN applications the power supply is limited, so preserving the consumed energy of the network is a challenge that must be considered when developing a routing protocol for WSNs.

A comprehensive survey of the routing protocols for WSNs can be found in \cite{Jiang}. In general, these protocols can be categorized into two classes according to the node's participating style: flat protocols and clustering protocols. Those in \cite{Hedetniemi,Heinzelman99,Sohrabi,Intanagonwiwat} belong to the first class.The second class can be also categorized into two subclasses: the clustering algorithms applied in homogeneous networks are called homogeneous schemes, where all nodes have the same initial energy and the clustering algorithms applied in heterogeneous networks are referred to as heterogeneous clustering schemes, where all the nodes of the sensor network are equipped with different amounts of energy.

Many homogeneous clustering algorithms exist in literature such as LEACH \cite{Heinzelman00}, PEGASIS \cite{PEGASIS}, HEED \cite{HEED} and RE-LEACH \cite{RELEACH}. Low-Energy Adaptive Clustering Hierarchy (LEACH), which is one of the most fundamental protocol frameworks in the literature, utilizes randomized rotation of the Cluster-Heads (CHs) to uniformly distribute the energy budget across the network. The sensor nodes are grouped into several clusters and in each cluster, one of the sensor nodes is selected to be CH. Each node will transmit its data to its own CH which forwards the sensed data to the BS finally. Both communications between sensor nodes and CH and that between CHs and the BS are direct, single-hop transmission. Based on the framework of  LEACH, several protocols are proposed in the open literature.
In \cite{PEGASIS}, a scheme called Power-Efficient GAthering in Sensor Information System (PEGASIS) is proposed. In this system, each node communicates only with a close neighbor and takes turns transmitting to the BS, thus reducing the amount of energy spent per round. In \cite{PEGASIS}, nodes will be organized to form a chain, which can be computed by each node or by the base station. The requirement of global knowledge of the network topology makes this method difficult to implement.
In\cite{HEED}, HEED is a distributed clustering algorithm, which selects the cluster-heads stochastically. The election probability of each node is correlative to the residual energy. But in heterogeneous environments, the low-energy nodes could own larger election probability than the high-energy nodes in HEED.

WSNs are more possibly heterogeneous networks than homogeneous ones. Thus, the protocols should be fit for the characteristic of heterogeneous WSNs. Many heterogeneous clustering algorithms exist in literature such as SEP\cite{SEP}, M-LEACH\cite{M-LEACH} , EECS\cite{EECS}, LEACH-B\cite{LEACH-B}, DEEC\cite{DEEC} and SDEEC\cite{SDEEC}.
The EECS\cite{EECS} protocol elects the cluster-heads with more residual energy through local radio communication. In cluster formation phase, EECS considers the tradeoff of energy expenditure between nodes to the cluster-heads and the cluster-heads to the base station. But on the other hand, it increases the requirement of global knowledge about the distances between the cluster-heads and the base station.
The EEHC \cite{EEHC} protocol is developed for the 3-level heterogeneous networks, which include three types of nodes according to the initial energy, i.e., the super nodes, the advance nodes and the normal nodes. The rotating epoch and election probability is directly correlated with only the initial energy of nodes. EEHC performs poorly when heterogeneity is a result of operation of the sensor network.

In this paper, we also focus on the design of power efficient network layer solutions. Our work is inspired by the previous approaches, but it differs by designing the protocol with the integration of the cross-layer design principle which is proven to be a pertinent method to meet the challenges of power-constrained WSNs. The EAMS protocol assigns different epoch of  being a cluster-head to each node according to the initial and residual energy. A novel clustering-based routing protocol proposed in this paper improve the effective life span of the WSNs with a limited energy supply.

EASM is an energy-aware scheme clustering used in heterogeneous wireless sensor networks. In witch, every sensor node independently elects itself as a cluster-head based on its initial energy and residual energy. To control the energy expenditure of nodes by means of adaptive approach, our algorithm use the orientation to BS to transmit the sensing data, and doesn't require any global knowledge of energy at every election round.

\subsection{Heterogeneous WSN model}
\begin{figure}[ht]
    \center
      \includegraphics[width=8cm, height= 12cm]{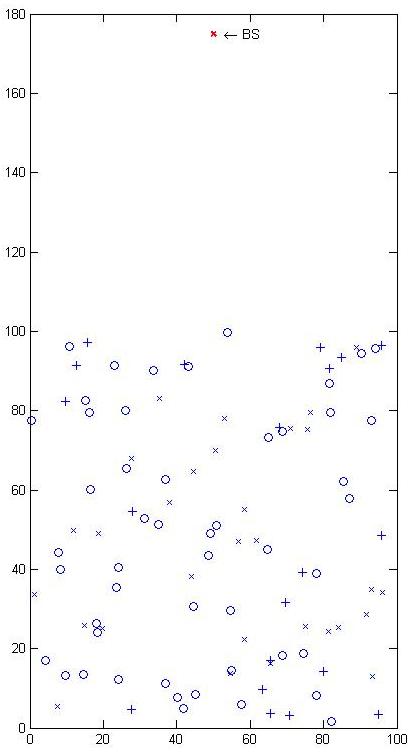}
      \caption[text1]{100 nodes randomly deployed in the network \\\hspace{\linewidth} o: normal node; x: advanced node; +: super node}
      \label{fig:1}
\end{figure}

In this study, we describe the network model. Assume that there are N sensor nodes, which are uniformly dispersed within an M $\times$ M square region (Fig.~\ref{fig:1}).

The nodes always have data to transmit to a base station, which is often far from the sensing area. The network is organized into a clustering hierarchy, and the cluster-heads execute data aggregation to reduce redundant data produced by the sensor nodes within the clusters.

We consider the heterogeneous networks with nodes heterogeneous
in their initial amount of energy. We assume there are
three types of sensor nodes, i.e.,the super nodes, the advanced nodes and the normal
nodes\cite{EEHC}.
Note $E_{0}$ the initial energy of the normal nodes, and
$ m $ the fraction of the total nodes N, and $m_{0}$ is the percentage  of the total
 number of nodes $ m $ which are equipped with $\beta$ times more energy than the normal
 nodes, we call these nodes as super nodes. The rest $N\times(1 - m_{0}) $ nodes are
 equipped with $\alpha$ times more energy than the normal nodes, we call these nodes
 as advanced nodes and the remaining $N\times(1 - m) $ as normal nodes.

Thus there are $N\times m\times(1-m_{0})$ advanced nodes equipped with initial energy
of $E_{0}\times(1 + \alpha)$, $N \times m\times m_{0}$ super
nodes equipped with initial energy of $E_{0}\times(1 + \beta)$
and $(1 - m)\times N$ normal nodes equipped with initial energy
of $E_{0}$.

The total initial energy of the three-level heterogeneous networks is given by:

\begin{eqnarray}
E_{tot}&=& N\times m\times(1-m_{0})\times E_{0}\times(1 + \alpha)          \\ \nonumber
       &+& N \times m\times m_{0}\times E_{0}\times(1 + \beta) + (1 - m)\times N\times E_{0} \\ \nonumber
       &=& N\times E_{0}\times(1+m\times(\alpha + m_{0}\times(\beta-\alpha)))
\label{eq:01}
\end{eqnarray}

\begin{figure}
    \center
      \includegraphics[width=8cm,height= 12cm]{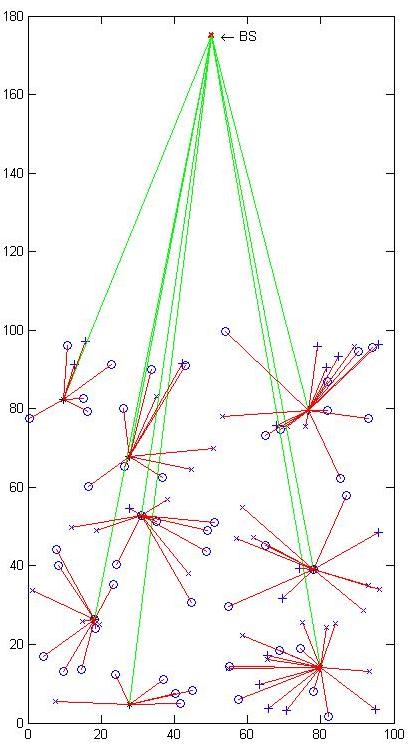}
      \caption{Dynamic cluster structure by EASM algorithm}
      \label{fig:2}
\end{figure}
\begin{figure}[!h]
    \center
      \includegraphics[width=8cm, height=4cm]{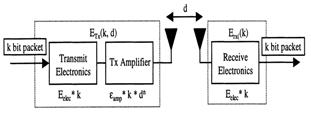}
      \caption{Radio Energy Dissipation Model}
      \label{fig:3}
\end{figure}

The cluster-heads (Fig. ~\ref{fig:2}) transmit the aggregated data to the BS directly. We assume that the nodes are stationary as supposed in \cite{Heinzelman02}.
More interestingly,  a similar energy model as proposed in \cite{Heinzelman02}is used in this study. According to the radio energy dissipation model illustrated in (Fig.~\ref{fig:3}), and in order to achieve an acceptable Signal-to-Noise Ratio (SNR) in transmitting an L-bit message over a distance d, the energy expended by the radio is given by :
\begin{equation}
E_{Tx}(l,d) = \left\{
\begin{array}{rl}
 lE_{elec}+l\epsilon_{fs}d^{2} , d < d_0
 \\
lE_{elec}+l\epsilon_{mp}d^{4} , d \geq d_0%
\end{array}%
\right.
\label{eq:02}
\end{equation}

Where $E_{elec}$ is the energy dissipated per bit to run the transmitter $E_{Tx}$ or the receiver $E_{Rx}$ circuit, and $\epsilon_{fs}$ and $\epsilon_{mp}$ depend on the transmitter amplifier model used and d is the distance between the sender and the receiver.

In most WSN applications the power supply is limited, so preserving the consumed energy of the network is a challenge that must be considered when developing a routing protocol for WSNs. In the next section, we describe the EASM algorithm in details.

\section{Explanation of the proposed protocol: (EASM)}
\label{sec:3}

We assume a network with N nodes uniformly deployed within M$\times$ M square region, the network topology remains stagnant over time and the BS location is known. In EASM, a new optimal probability threshold is introduced, where each node i uses to determine whether itself to become a cluster-head in each round r , given as follows:

\begin{equation}
T(s_{i}) = \left\{
\begin{array}{rl}
 \frac{p_{i}}{1-p_{i}(r mod \frac{1}{p_{i}} )}\times\frac{E_{residual} (r_{i})}{E_{initial}(i)} \ ,  \ if \  s_{i}  \ \ \epsilon  \ G
  \\
  \\
  0 \ \ \ \ \  \  \ \ \ \      , \ \ otherwise%
\end{array}%
\right.
\label{eq:03}
\end{equation}

Where $E_{residual} (r_{i})$, $E_{initial}(i)$ are the residual and the initial energy respectively. $p_{i}$ is the cluster-head probability and  $r_{i}$ is the number of consecutive rounds in which a node has not been cluster-head within an epoch.

When $r_{i}$ reaches the value  $1/p_{opt}$ the threshold  $T(i)_{opt}$  is reset to the value it had before the inclusion of the remaining energy into the threshold-equation (~\ref{eq:03}).

Also, the probabilities for normal, advanced and super nodes are defined as follow:
\begin{eqnarray}
p_{n} &=& \frac{p_{opt}}{(1+m\times(\alpha + m_{0}\times(\beta-\alpha)))}
\label{eq:04}
\end{eqnarray}
\begin{eqnarray}
p_{a} &=& \frac{p_{opt}}{(1+m\times(\alpha + m_{0}\times(\beta-\alpha)))}\times(1+\alpha)
\label{eq:05}
\end{eqnarray}
\begin{eqnarray}
p_{s} &=& \frac{p_{opt}}{(1+m\times(\alpha + m_{0}\times(\beta-\alpha)))}\times(1+\beta)
\label{eq:06}
\end{eqnarray}
where $ m $ is the fraction of the total nodes N, and $m_{0}$ is the percentage  of the total
number of nodes $ m $ which are equipped with $\beta$ times more energy than the normal
nodes; The rest nodes are  equipped with $\alpha$ times more energy than the normal nodes.

In each round r, when node i finds it is eligible to be a cluster head, it will choose a random number between 0 and 1. If the number is less than threshold $T_{opt}(i)$, the node i becomes a cluster head during the current round.

Each node that has elected itself a cluster-head for the current round broadcasts an advertisement message to the rest of the nodes. For this "cluster-head-advertisement" phase, the cluster-heads use a CSMA MAC protocol, and all cluster-heads transmit their advertisement using the same transmitted energy. The non-cluster-head nodes must keep their receivers on during this phase of set-up to hear the advertisements of all the cluster-head nodes. After this phase is complete, each non-cluster-head node decides the cluster to which it will belong for this round. This decision is based on the received signal strength of the advertisement. Assuming symmetric propagation channels, the cluster-head advertisement heard with the largest signal strength is the cluster-head to whom the minimum amount of transmitted energy is needed for communication. In the case of ties, a random cluster-head is chosen \cite{Heinzelman00}.

Each non cluster head node communicates its data during its allocated transmission time (TDMA) to its own cluster head. After that, each non cluster head can turn on the sleep mode. The cluster head node must keep its receiver on in order to receive all the data from the nodes in the cluster.

When all the data is received, the cluster head node performs signal processing functions to compress the data into a single signal. When this phase is completed, each cluster head can send the aggregated data to the BS.

%%%%%%%%%%tableau===========================
\begin{table}
\caption{Radio parameters used in our simulations}
\label{tab:1}       % Give a unique label
%
% Follow this input for your own table layout
%
\begin{tabular}{p{3cm}p{5cm}}
\hline\noalign{\smallskip}
Parameter  &\vline  Value  \\
\hline\noalign{\smallskip}\noalign{\smallskip}
$E_{elec}$ &  $5 nJ/bit$\\
$\epsilon_{fs}$ &  $10pJ/bit/m^2$\\
$\epsilon_{mp}$ &  $0.0013pJ/bit/m^4$\\
$E_0$ &  $0.5 J$\\
$E_{DA}$ &  $5 nJ/bit/message$\\
$d_{0}$ &  $70 m$\\
Message size &  4000 bits\\
$p_{opt}$ &  $0.1$\\
\noalign{\smallskip}\hline\noalign{\smallskip}
\end{tabular}
\end{table}
%%%%%%%%%%tableau==========================

\section{Simulation}
\label{sec:4}
In this section, we evaluate the performance of EASM protocol. We consider a WSN with N = 100 nodes randomly distributed in a 100m $\times$ 100m sensing area. We assume the BS is far away from the sensing region and placed at location$(x = 50;y = 175)$. The nodes in the network are divided in three  heterogenous energy levels and are energy-constrained.

To compare the performance of EASM with other protocols, The radio parameters used in our simulations are shown in TABLE~\ref{tab:1}. We assume that all nodes know their location coordinates. We will consider the following scenarios and examine several performance measures.

After deployment of WSN, the nodes consume energy during the course of the WSN life span. In fact, energy is
removed whenever a node transmits or receives data and whenever it performs data aggregation. Once a node runs out of energy, it is considered dead and can no longer transmit or receive data.

Firstly, we run simulation for our proposed protocol EASM to detect the round when the first node dies and compare the results to LEACH and EEHC protocols under two kinds of 3-level heterogeneous networks .
\begin{figure}[!h]
    \center
      \includegraphics[width=8cm,height=7cm]{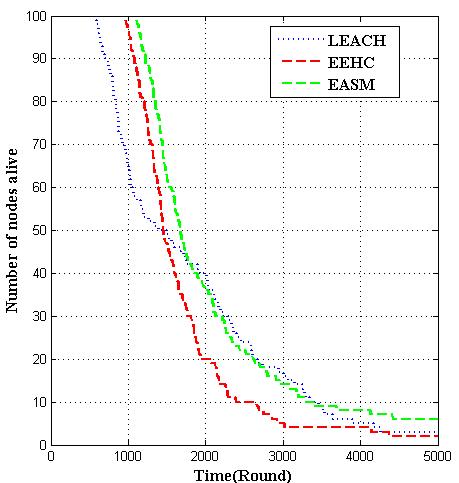}
      \caption{Number of nodes alive over time.($\alpha=1.5$,$m =0.5$,$\beta=3$ and $m_{0}=0.4$)}
      \label{fig:4}
\end{figure}
Figure Fig.~\ref{fig:4} shows the results of the case with $\alpha=1.5$, $m =0.5$, $\beta=3$ and $m_{0}=0.4$. It is obvious that the stable time of EASM is prolonged compared to that of LEACH and EEHC.

\begin{figure}[!h]
    \center
      \includegraphics[width=8cm,height=7cm]{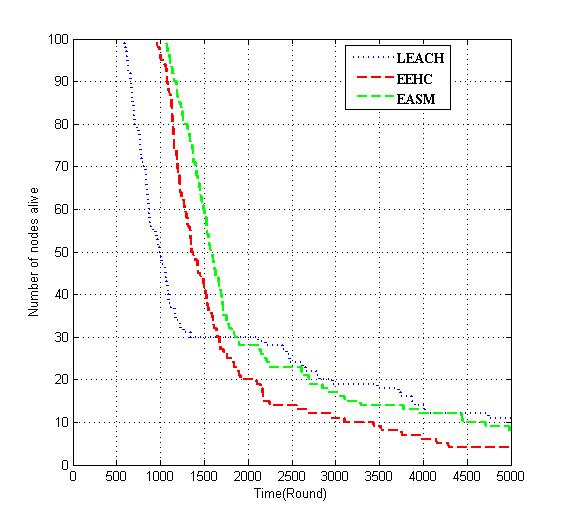}
      \caption{Number of nodes alive over time.($\alpha=2$,$m =0.3$,$\beta=5$ and $m_{0}=0.6$)}
      \label{fig:5}
\end{figure}

Second, we run simulation for our proposed protocol EASM to compute the round when the first node dies when $\alpha=2$,$m =0.3$,$\beta=5$ and $m_{0}=0.6$, and compare the results to LEACH and EEHC protocols. Fig.\ref{fig:5} shows the number of rounds when the first node dies.

For EEHC, the stability period of EEHC is much longer than that of LEACH. Though  achieves the stability
period longer by about 37\% than LEACH (see Fig.\ref{fig:4} and \ref{fig:5}). This is
because EEHC is an energy-aware protocol, which elects the cluster-heads according to the residual energy of nodes. Being
also an energy-aware protocol, EASM outperforms other clustering protocols. In fact, EASM obtains 19\% more rounds than EEHC.

\begin{figure}[!h]
    \center
      \includegraphics[width=8cm]{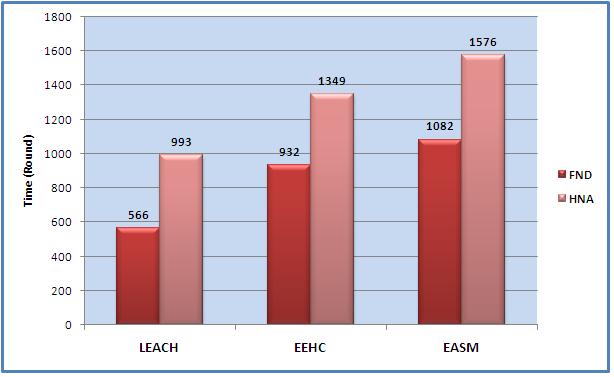}
      \caption{FND and HNA}
      \label{fig:6}
\end{figure}

Fig. \ref{fig:6} shows the comparison between all nodes in terms of FND and HNA. Obviously, we can remark that our protocol EASM
contains a larger period of stability time than LEACH and EEHC, which increases the efficiency of the network.
We notice the same results for HNA.

A longer stable time metric is important because it gives the end user reliable information of the sensing area, which
extend the network lifetime. This reliability is vital for sensitive applications like tracking fire in forests.

\begin{figure}[h]
    \center
      \includegraphics[width=8cm,height=7cm]{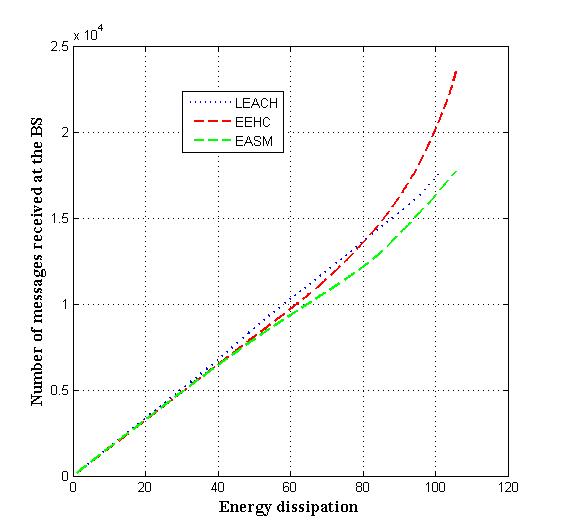}
      \caption{Number of message received at the BS over energy spent}
      \label{fig:7}
\end{figure}

Third, we run simulation for our proposed protocol EASM to compute the number of received messages at the BS over
energy dissipation and compare the results to LEACH and EEHC protocols. Fig.\ref{fig:7} shows that the
messages delivered by EASM to the BS are better than the others ones; this means that EASM is an energy-aware
adaptive clustering protocol.

\begin{figure}[h]
    \center
      \includegraphics[width=8cm,height=7cm]{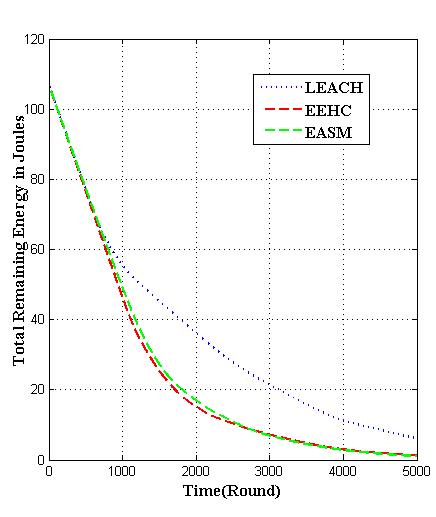}
      \caption{Total remaining energy over rounds under 3-level heterogeneity of LEACH, EEhC and EASM}
      \label{fig:8}
\end{figure}

Fig. \ref{fig:8} shows the remaining energy over time for all simulated protocols and it reveals that
EASM consumes less energy in comparison to the others, which helps to extend the network
life span.

According to the simulation results, we can obviously state that EASM is a more efficient protocol than LEACH and EEHC, and consequently can be considered as an energy-aware protocol.

\section{Conclusion}
\label{sec:5}
It has been explained in details that EASM is an energy-aware adaptive clustering protocol used in Multi-level heterogenous WSNs. To control the energy expenditure of nodes by means of adaptive approach, EASM uses new optimal probability threshold which takes the ratio of residual energy and initial energy into account. In order to increase more the EASM protocol performances, we implemented a dynamic way to distribute the spent energy more equitably between nodes. Thus, saves energy in a better way and consequently increases the life span of the WSNs

To sum up, we can say that the proposed algorithm EASM extends and outperforms better the performances of EEHC protocol.
%\section*{Acknowledgment}

%\input{reference}

\bibliographystyle{ieeetr}
\bibliography{aesm_edspia}

\textbf{Authors:}\\
\textbf{Mostafa SAADI :} Received the B.Sc. degree in Computer Sciences at the University
Hassan $2^{nd}$ , Faculty of Sciences Ain-Chook, Casablanca, Morocco, in 2003, and a
M.Sc. degree in Mathematical and Computer engineering at the University Chouaib
Doukkali, Faculty of Sciences, El Jadida (FSJ), Morocco, in 2009. He has been
working as a professor of Computer Sciences in high school since 2003, in Sidi Rahal
Beach, Morocco. Currently, he is working toward his Ph.D. at  FSJ. His current
research interests performance evaluation, analysis and simulation of Wireless Sensor
networks.

\textbf{Dr. Moulay Lahcen HASNAOUI:} Received his Ph.D in modelling and simulation
of semiconductor devices at the Paris-Sud University, France (1991-1995). He
worked as research associate in developing fuel cell at Department of Engineering
Physics, Polytechnic School, Montreal, Canada (1996-1996). He earned his
bachelor's degree in Computer Science from University of Montreal, Canada (1998-
2002). Self-employed as a software developer (2002-2004). He worked as research
assistant professor at Mathematics and Computer Science Department at the Faculty
of Sciences, MATIC Laboratory, El Jadida, Morocco,  between 2004-2011. He is working as research
assistant professor at Computer Sciences Department at the Faculty of Sciences Dhar Al Mahraz, Fez
(2011).

\textbf{Abderrahim BENI HSSANE:} Is a research and an assistant professor at Science
Faculty, Chouaîb Doukkali University, El Jadida, Morocco, since September 1994.
He got his B.Sc. degree in applied mathematics and  his Doctorate of High Study
Degree in computer science, respectively, in 1992 and 1997 from Mohamed V
University, Rabat, Morocco. His research interests focus on performance evaluation
in wireless networks.

\textbf{Said BENKIRANE:} Obtained his Certificate in telecommunications engineering at
the National Institute of Posts  and  Telecommunications, Rabat,  Morocco, in  2004,
and his M.Sc. degree in computer engineering and network from the University of
Sidi Mohammed Ben Abdellah Fez, Morocco in 2006. He has been working as
professor of Computer Sciences in high school since 2007, in El Jadida, Morocco.
He is a member of a research group e-NGN (e-Next Generation Networks) for Africa
and Middle East. Currently, he is pursuing his Ph.D at the Faculty of Sciences,
Chouaib Doukkali University, El Jadida, Morocco. His main research areas include wireless and mobile
computing and mobile telecommunications systems.

\end{document}